\documentclass[amssymb,prb,aps,twocolumn,amsmath,showpacs]{revtex4}

\usepackage[dvips]{graphicx}
\usepackage{amstext}

\begin{document}

\title{Critical Current Behavior in Josephson Junctions with the Weak Ferromagnet PdNi}
\author{Trupti S. Khaire, W. P. Pratt, Jr., Norman O. Birge}
\email{birge@pa.msu.edu}
\affiliation{Department of Physics and
Astronomy, Michigan State University, East Lansing, Michigan
48824-2320, USA}
\date{\today}

\begin{abstract}
We have studied the variation of critical current in
Superconductor/Ferromagnet/Superconductor (S/F/S) Josephson
Junctions as a function of ferromagnet thickness using a
weakly-ferromagnetic alloy Pd$_{82}$Ni$_{12}$. Measurements were
performed for the thickness range 32 to 100 nm, over which the
critical current density decreases by five orders of magnitude.
The critical current density oscillates with a period of $\pi
\times 4.1\pm 0.1$\,nm, and decays over a characteristic length of
$8.0\pm 0.5$\,nm. There is no evidence of a crossover to a slower
decay, which might indicate the presence of long-range
spin-triplet pair correlations. We discuss possible reasons for
their absence, including the possibility of strong spin flip
scattering in PdNi.
\end{abstract}

\pacs{74.50.+r, 74.45.+c, 75.30.Gw, 74.20.Rp} \maketitle

\section{Introduction}

When a superconducting (S) metal is placed in contact with a
normal (N) metal, the properties of both metals are modified near
the S/N interface.  The resulting superconducting proximity effect
was widely studied in the 1960's,\cite{deGennes} and again in the
1990's.\cite{LambertRaimondi,Pannetier:2000} When the normal metal
is replaced by a ferromagnetic (F) metal, the resulting physics is
extremely rich, due to the very different order parameters in the
two metals. There has been sustained interest in S/F systems over
at least the past decade.\cite{BuzdinReview,PokrovskyReview} Many
new phenomena have been observed, including but not limited to:
oscillations in the critical temperature of S/F bilayers as a
function of F layer thickness,\cite{Jiang:95} variations in the
critical temperature of F/S/F trilayers as a function of the
relative magnetization direction of the two F layers,\cite{Gu:02}
and oscillations in the critical current of S/F/S Josephson
junctions.\cite{Oboznov:06} The oscillatory behaviors are a direct
result of the exchange splitting of the spin-up and spin-down
bands in the ferromagnet, which produce a momentum shift between
the up- and down-spin electrons of a Cooper pair that ``leaks"
from S into F.\cite{Demler} In the clean limit, the spatial period
of the oscillations is governed by the exchange length,
$\xi_F=\hbar v_F/2E_{ex}$, where $v_F$ and $E_{ex}$ are the Fermi
velocity and exchange energy of the F material, respectively.

Unlike the S/N proximity effect, which extends to distances of
order 1 $\mu m$ at sufficiently low temperature, the novel
phenomena associated with the S/F proximity effect persist only
over very short distances -- limited either by $\xi_F$, by the
mean free path, $l_e$, or by their geometric mean, the dirty-limit
exchange length $\xi_F^*=\sqrt(\hbar D/E_{ex})$, where $D=v_F
l_e/3$ is the diffusion constant.  These distance scales,
characterizing the oscillation and decay of superconducting
correlations, tend to be extremely short in strong ferromagnets,
and only moderately longer in the weakly-ferromagnetic alloys
preferred by some groups.  For example, oscillations in the
critical current in Nb/Co/Nb Josephson junctions have been
observed with a period of 1.0 nm and a decay constant of 3.0
nm.\cite{Robinson:06} In Nb/Cu$_{47}$Ni$_{53}$/Nb alloy junctions,
the observed period and decay constants were 11.0 nm and 1.3 nm,
respectively.\cite{Oboznov:06}

In this context, it was very exciting when two theoretical groups
predicted that a new form of superconducting order, with
spin-triplet pairing, could be induced by certain types of
magnetic inhomogeneity in S/F systems consisting of conventional
spin-singlet superconducting
materials.\cite{Shekhter:01,Bergeret:01,Bergeret:03-1,Bergeret:03-2}
Cooper pair correlations with spin-triplet symmetry are not
subject to the exchange field of the ferromagnetic, since both
electrons in the pair enter the same spin band in the ferromagnet.
As a result, proximity effects due to such correlations should
persist over long distances in a ferromagnetic, limited either by
the temperature or by spin-flip and spin-orbit scattering.
Furthermore, the spin-triplet correlations predicted to occur in
S/F systems are not of same type discovered recently in materials
such as SrRuO$_4$.\cite{Eremin:04} In SrRuO$_4$, the spin-triplet
Cooper pairs satisfy the Spin-Statistics Theorem by having odd
orbital angular momentum. In contrast, the spin-triplet
correlations predicted to occur in S/F systems have even orbital
angular momentum, and satisfy the Spin-Statistics Theorem by
virtue of being odd in time or frequency.\cite{BergeretReview}

Experimental confirmation of the presence of spin-triplet
correlations is not easy. In retrospect, several theorists have
suggested their role in old experiments performed on mesoscopic
S/F hybrid samples, where the data were interpreted in terms of a
long-range superconducting proximity
effect.\cite{Giordano,Petrashov,Courtois}  More recently, there
has been one report of a Josephson current in S/F/S structures
using CrO$_2$ as the F material,\cite{Keizer} where the distance
between the S layers was very long (several hundred nm), so that
the conventional spin-singlet supercurrent should be exponentially
suppressed.  In a different work,\cite{Sosnin:06} phase-coherent
oscillations were observed in the normal resistance of a Ho wire
connected to two superconducting electrodes, again separated by a
distance too large to support spin-singlet correlations.  In both
cases, the data were interpreted as being due to spin-triplet
superconducting correlations in the ferromagnetic material. While
these pioneering experiments are highly suggestive and
tantalizing, the first suffers from large sample-to-sample
fluctuations in the magnitude of the observed
supercurrent,\cite{Keizer} while both lack direct evidence for
spin-triplet correlations.

The goal of this work is to study S/F/S Josephson junctions where
the thickness of the F layer is increased systematically, from a
thin regime where the supercurrent is likely to be dominated by
spin-singlet correlations, to a thick regime where the smaller but
longer-range spin-triplet supercurrent takes over.  We chose Nb as
the S material because its large critical temperature allows us to
make measurements at 4.2 K, and hence measure a large number of
samples.  For the F material, we chose Pd$_{1-x}$Ni$_{x}$ alloy
with a Ni concentration of 12 atomic \%, a material studied
extensively by the group of M. Aprili\cite{Kontos:01,Kontos:02}
and others,\cite{Matsuda,Cirillo,Cirillo:02,Cirillo:03,Bauer} but
only with PdNi thicknesses less than 15 nm.  Our choice of this
particular weakly-ferromagnetic alloy was based on two
considerations: 1) Using a weakly-ferromagnetic material allows us
to increase the thickness of the F layer without introducing an
overwhelming amount of intrinsic magnetic flux inside the
junctions.  (This issue will be discussed further in Section III
below.)  2) Some weakly-ferromagnetic alloys suffer from strong
spin-orbit and/or spin-flip scattering, which are likely to
destroy both spin-singlet and spin-triplet correlations. For
example, Ryazanov and co-workers\cite{Ryazanov:01,Oboznov:06} have
found that the Josephson current in S/F/S junctions using
Cu$_{1-x}$Ni$_{x}$ alloy with x = 53 at.\% decreases exponentially
over a length scale of only 1.3 nm. Moreover, that length scale is
much shorter than the one characterizing the critical current
oscillations, a fact that implicates strong spin-flip scattering
in CuNi alloy.\cite{Demler,Bergeret:03-2,Faure:06}  In contrast,
while Kontos \textit{et al.}\cite{Kontos:02} found that the
critical current in S/F/S junctions made with PdNi alloy decays
over a length scale only slightly longer, 2.8 nm, they found that
the oscillations and decay are governed by the \textit{same}
length scale, which may imply that spin-flip and spin-orbit
scattering are weak in this material.

The paper is organized as follows.  In Section II we discuss
sample fabrication methods and characterization of the PdNi alloy.
Section III discusses the characterization of our S/F/S Josephson
junctions, with particular attention to the magnetic-field
dependence of the critical current (the so-called ``Fraunhofer
pattern").  Section IV presents the main results of the paper,
namely the critical current vs. PdNi thickness of our S/F/S
Josephson junctions.  Section V discusses the various theoretical
works on S/F/S junctions, and the physical parameters that one can
extract from fitting theoretical formulas to the data. Our
interpretation of the results is presented in Section VI. Finally,
we conclude with suggestions for future directions.

\section{Experimental}
\subsection{Sample Fabrication}

Substrates were silicon chips of dimension $12.7 \times 12.7$ mm.
Preparation for deposition was performed in a cleanroom, to
minimize the presence of dust particles which could lead to shorts
in the Josephson junctions. A multilayer consisting of
Nb(150)/PdNi(d$_{PdNi}$)/Nb(25)/Au(15) (with all thicknesses in
nm) was deposited using magnetically-enhanced triode dc sputtering
in an Ar plasma pressure of 2.5 mTorr after obtaining a base
pressure of 2 x $10^{-8}$ Torr or better. A mechanical shadow mask
was used to create the multilayer strip of size $0.16 \times 10$
mm$^2$. The thin Au protective layer prevents oxidation of the top
Nb layer during further processing steps. The PdNi thickness,
$d_{PdNi}$ was varied from 32.5 nm to 100 nm, typically in 5 nm
steps.  A subset of samples was fabricated with more
closely-spaced thicknesses in the range from 58 nm to 75 nm, in
order to demonstrate the minima in critical current indicative of
the $0-\pi$ transitions.

The multilayer was patterned using photolithography to create
circular photoresist pillars with diameters of 10, 20, 40, and 80
$\mu m$ on each substrate. Care was taken to ensure the presence
of undercut in the resist profile. Either a trilayer
photolithography consisting of two photoresist layers separated by
a thin metallic layer or a single layer photolithography using
chlorobenzene yielded large reliable undercuts in the resist
profile. The large variation in area allowed us to have a large
dynamic range for the critical current measurements.  In practice,
however, the success rate of the 80 $\mu m$ pillars was very low,
possibly due to the presence of dust particles during one of the
fabrication steps taking place outside the cleanroom.

The photoresist pillars acted as a mask to protect the multilayer
below them while the rest was ion milled. The multilayer was
milled down to the middle of the ferromagnetic layer, thus
completely removing the top Nb layer yet not exposing the bottom
Nb layer, to prevent the possibility of back-sputtered Nb
depositing on the sides of the circular pillars. Nearly 200 nm of
$SiO_{x}$ was then deposited to insulate the bottom Nb from the
top Nb leads. Lift off of the photoresist pillars was then done
using Remover PG. This was followed by a slight ion milling of the
top Au layer to ensure a clean interface. The top Nb lead of
thickness 150 nm was then sputtered in the end. The thin Au layer
becomes superconducting due to the proximity effect, as it is
sandwiched between two Nb layers.  A schematic of a complete
Josephson junction is shown in Fig. 1.

\begin{figure}[tbh]
\begin{center}
\includegraphics[width=3.2in]{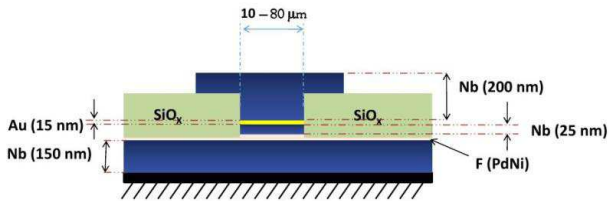}
\end{center}
\caption{(color online). Schematic of a ferromagnetic Josephson
junction.} \label{Schematic}
\end{figure}

\subsection{Characterization of PdNi Alloy}

\begin{figure}[tbh]
\begin{center}
\includegraphics[width=3.in]{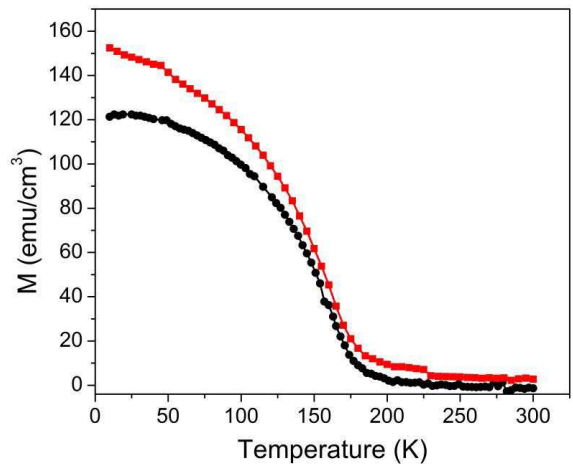}
\end{center}
\caption{(color online). Magnetization vs. temperature for a
100-nm thick PdNi film, grown on 150 nm of Nb to have identical
crystalline properties as our Josephson junction samples. The film
was first cooled in zero field to 10 K, then magnetized by
applying an in-plane field (black circles) or out-of-plane field
(red squares) of 5 kG, then the magnetic moment was measured in
field while heating to room temperature. The cooling curves (not
shown) were also measured, and were found to follow the same
curves as during heating except for occasional jumps seen for the
out-of-plane field case.} \label{MvsT}
\end{figure}

The Ni concentration of our PdNi alloy was estimated by three
different methods.  Energy dispersive X-ray analysis (EDAX),
performed on a $1.5\mu m$ thick PdNi film, yielded a Ni
concentration of $12\pm0.5\%$.  (A thick film was used for this
measurement to increase the signal-to-noise ratio for the Ni-K
peak. Similar measurements performed on 200-nm thick PdNi film
yielded a similar concentration value provided the signal was
accumulated for long enough times.)  To corroborate this value for
the concentration, the magnetization $M$ vs. temperature of a
100-nm thick PdNi film was measured using a Quantum Design SQUID
magnetometer (see Fig. \ref{MvsT}). A clear change in the slope is
seen around the Curie temperature of about 175 K, independent of
whether the magnetic field is applied in-plane or out-of-plane.
This Curie temperature corresponds to a Ni concentration of
$12\%$, according to earlier work by Beille,\cite{Beille} cited by
Kontos.\cite{Kontos-thesis}

\begin{figure}[tbh]
\begin{center}
\includegraphics[width=3.in]{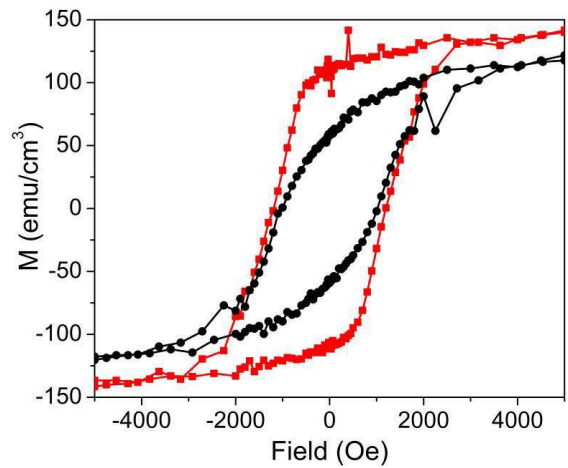}
\end{center}
\caption{(color online). Magnetization vs. in-plane field (black
circles) and out-of-plane field (red squares) for the same 100-nm
thick PdNi film shown in Fig. \ref{MvsT}, measured at 10 K.}
\label{MvsH}
\end{figure}

Fig. (\ref{MvsH}) shows $M$ vs. $H$ at $T=10$K for the same 100-nm
thick PdNi film. The magnetization curve is more rounded, and has
smaller remanent magnetization, when the field is applied
in-plane, indicating that the magnetic anisotropy of PdNi films is
out-of-plane. Similar measurements on PdNi films of thickness 30
and 60 nm also indicate out-of-plane anisotropy, but somewhat less
pronounced than in the 100-nm film.  The out-of-plane anisotropy
of PdNi surprised us initially, because the strong shape
anisotropy of thin films usually dominates over magnetocrystalline
anisotropy, resulting in overall in-plane anisotropy. Out-of-plane
anisotropy has been observed in several materials, however,
including CuNi alloy.\cite{Ruotolo,Veshchunov} Recently, we
learned that Aprili has also observed out-of-plane anisotropy in
PdNi films.\cite{Aprili_private} The implications of PdNi's
out-of-plane anisotropy on our work will be discussed further at
the end of this paper.

The saturation magnetization of PdNi measured out-of-plane at 10 K
is $M_{sat}= 150$ emu/cm$^3$ = 0.23 $\mu_B$/atom (see Fig.
\ref{MvsH}).  According to refs.
[\onlinecite{Beille,Kontos-thesis}], this corresponds to a Ni
concentration of nearly $12\%$, in agreement with the
determination from the EDAX measurements and Curie temperature.

\subsection{Ic measurement}

The normal-state resistances of our Josephson junction pillars
vary from 2.4 $\mu \Omega$ to 152 $\mu \Omega$, depending mostly
on the pillar area, and, to a lesser extent, on the PdNi
thickness. Resistances in this range require an extremely
sensitive low noise measurement technique which is provided by
using a superconducting quantum interference device (SQUID) as a
null detector in a current-comparator circuit.\cite{EdmundsPratt}
All the four probe measurements were performed at 4.2 K by dipping
the probe into a liquid Helium dewar equipped with a cryoperm
shield. Each chip had Josephson junctions of diameters 10, 20, 40,
and 80 $\mu m$.  All measurements reported here were performed on
junctions having Josephson penetration depth, $\lambda_J =
(\Phi_0/(2 \pi \mu_0 J_c (d_F+2 \lambda_L))^{1/2}$ larger than
one-quarter of the junction diameter $w$.  ($\Phi_0 = h/2e$ is the
superconducting flux quantum, $J_c$ is the critical current
density, and $\lambda_L$ is the London penetration depth, equal to
about $86$ nm in our sputtered Nb.) This ensures uniform current
density in the Josephson junction.\cite{Barone-Paterno} If
$\lambda_J \ll w$, then the flux is screened from the center of
the junction (Meissner effect) and the effect of the self field
cannot be neglected.

\section{Characterization of Josephson Junctions}

\begin{figure}[tbh]
\begin{center}
\includegraphics[width=3.in]{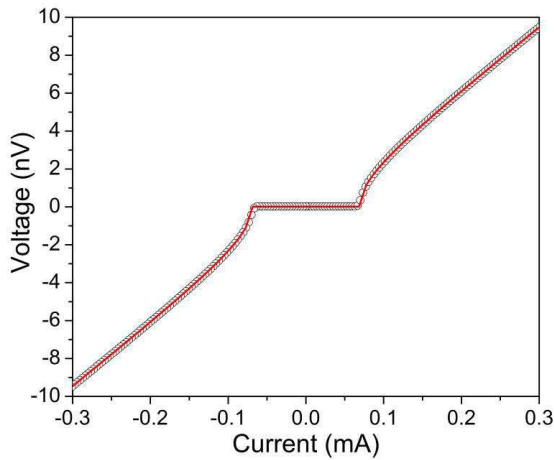}
\end{center}
\caption{(color online). Voltage vs. current for a Josephson
junction with diameter 20 $\mu$m and $d_{PdNi}$= 62 nm and
$H_{ext}$ = -19 Oe. The red solid line is a fit to Eqn.
\ref{I_vs_V}} \label{IvsV}
\end{figure}

Fig. \ref{IvsV} shows a I-V relation typical for our S/F/S
Josephson junctions.  The curve follows the standard form for
large-area, overdamped junctions:
\begin{equation}\label{I_vs_V}
V(I) = \frac{I}{|I|} R_N \textrm{Re}[(I^2-I_c^2)^{1/2}]
\end{equation}
Occasionally, we find that the I-V curves are shifted
horizontally, so that the critical current is not exactly the same
in the positive and negative current directions.  In such cases,
we average the critical currents in the two current directions.

One of the best ways to characterize Josephson junctions is to
observe the modulation of critical current as a function of
magnetic field $H_{ext}$ applied perpendicular to the current flow
direction in the junction. In nonmagnetic square junctions, the
pattern so obtained is called the Fraunhofer pattern, due to its
similarity to the pattern produced in single-slit diffraction of
light. Observation of a good Fraunhofer pattern demonstrates that
the supercurrent is uniform across the junction area, and that
there are no short circuits in the surrounding SiO$_x$ insulator.

In nonmagnetic Josephson junctions with circular cross section and
negligible screening ($\lambda_J > w$), the magnetic-field
dependence of the critical current is given by
\begin{equation}\label{ConvJJ}
I(\Phi)=I_C(0)\frac{2\times
J_{1}(\frac{\pi\Phi}{\Phi_{0}})}{(\frac{\pi\Phi}{\Phi_{0}})},
\end{equation}
where $I_{c}(0)$ is the critical current in the absence of
magnetic field, $J_{1}$ is the Bessel function of the first kind
of order 1, and $\Phi=H_{ext}(2\lambda_L + d)w$ is the magnetic
flux penetrating the middle of the Josephson junction, with
$\lambda_L$ the London penetration depth, $w$ the diameter of the
circular junction and $d$ the thickness of the barrier. This
pattern is called an ``Airy pattern." The pattern is qualitatively
similar to the Fraunhofer diffraction pattern, but the first
minima are spaced more widely apart in field than the subsequent
minima.

In samples containing a ferromagnetic barrier, one must include
the intrinsic flux of the magnetic barrier on the Fraunhofer
pattern.  If the magnetization $M$ is uniform throughout the
junction, the intrinsic magnetic flux is given by $\Phi_{F}=\mu_0
M d_{F} w$, with $d_F$ the F-layer thickness and $w$ the cross
section width (equal to the diameter for circular junctions). In
that case, the total magnetic flux through the F-layer is given by
\begin{equation}\label{Tot-mag-flux}
\Phi_{tot}=\mu_0 M d_{F} w + H_{ext} (2\lambda_L+d_{F})w
\end{equation}

\begin{figure}[tbh]
\begin{center}
\includegraphics[width=3.2in]{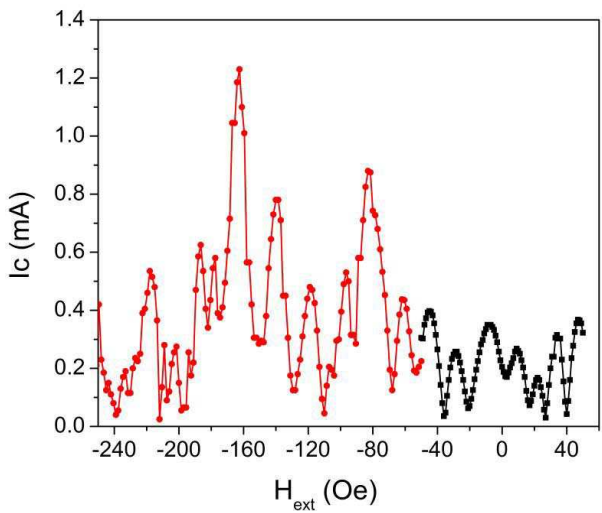}
\end{center}
\caption{(color online).  Critical current vs. in-plane magnetic
field for a Nb/Ni/Nb circular Josephson junction of diameter 10
$\mu$m, with $d_{Ni}=11$nm.  The black points (squares) were measured in the
virgin state, whereas the red points (circles) were measured after
magnetizing the sample in an external field of +1 kOe.  The random
pattern arises due to the intrinsic magnetic flux of the complex
domain structure of the Ni layer.} \label{FraunNickel}
\end{figure}

In macroscopic samples the magnetization breaks into domains, and
Eqn. (\ref{Tot-mag-flux}) is not valid.  Instead, one must
integrate the current density across the area of the junction,
taking into account the spatial dependence of the magnetic vector
potential $\vec{A}$ due to the domains. In the Coulomb gauge, one
takes $\vec{A}$ pointing along the current direction (z).  The
gauge-invariant phase difference across the junction includes a
term proportional to the line integral of $\vec{A}$ from deep
inside one superconducting contact to deep inside the
other.\cite{Levy}  The resulting expression for the supercurrent
is:
\begin{equation}\label{Shift}
I_s=\int J(x,y) dx dy
\end{equation}
where
\begin{equation}\label{Nofield} J(x,y)=J_{c} \sin (\alpha +
\frac{e}{\hbar c}\int \textbf{A(x,y)}.\textbf{dl})
\end{equation}
The term containing the vector potential performs a random walk as
one moves across the sample, due to the domains pointing in random
directions. If the magnetic domains are very small and/or the
magnetization is very weak, then the vector potential term stays
near zero in all parts of the junction, and the critical current
is hardly affected.  If, however, the magnetic domains are large
and/or have large magnetization as in the case of strong
ferromagnet, the contribution to the phase due to the vector
potential deviates far from zero as it crosses even a single
domain, thus severely suppressing the critical current. This can
lead to complete destruction of the Fraunhofer pattern. This is
clearly seen in Fig. \ref{FraunNickel}, which shows data for an
S/F/S junction of diameter 10 $\mu m$, with an 11-nm thick Ni
layer. (Similar random-looking ``Fraunhofer patterns" have also
been seen by other groups studying S/F/S Josephson
junctions.\cite{Bourgeois:01}) In principle, a regular Fraunhofer
pattern can be recovered if the sample is completely magnetized,
by applying a magnetic field in the opposite direction to the
magnetization such that the total flux given by Eqn.
(\ref{Tot-mag-flux}) is zero.  In that case, one should observe a
regular Fraunhofer pattern shifted in field by an amount:

\begin{equation}\label{Shift}
H_{shift}=\frac{-\mu_0 M d_{F}}{(2\lambda_L+d_{F})}
\end{equation}

The above argument holds only if the coercive field of the
magnetic material is large enough so that the magnetization stays
nearly uniform even in the presence of the applied field,
$H_{shift}$, pointing in the opposite direction.  For the case of
the Ni sample shown in Fig. \ref{FraunNickel}, the largest peak in
the critical current vs. field after magnetization is found near
160 Oe, whereas the expected shift calculated from the known
saturation magnetization of Ni is about 207 Oe. The discrepancy is
caused by some rotation of the magnetization in the domains or
some domain wall motion as $H_{ext}$ approaches the coercive
field, which we measured to be approximately 180 Oe in a 9-nm Ni
film.

To avoid the distortion of the Fraunhofer pattern, there are
several options for the study of ferromagnetic Josephson
junctions: 1) Use ultrathin samples: Under this condition the flux
enclosed in the junction due to a single magnetic domain is much
less than one flux quantum, and one can safely ignore the
contribution to the flux from the magnetization. This option is
not available to us, because our goal requires us to work with
thick ferromagnetic layers.  In addition, thin magnetic films
often have magnetically "dead" layers on each side, which pose
additional problems for ultra thin samples; 2) Use samples with
ultra-small lateral dimensions to reduce the contribution to the
total magnetic flux from the magnetization, and to control the
domain structure: This method has been pursued by Blamire and
co-workers \cite{Robinson:07} and also by Strunk and
co-workers,\cite{Strunk} using strong ferromagnets. The
disadvantage of this approach is that it becomes less effective as
the thickness of the ferromagnetic layer is increased.  In
addition, it restricts the possibilities to introduce magnetic
inhomogeneities that are naturally present due to the domain
structure in devices of larger dimension, and which may be crucial
for inducing the predicted spin-triplet superconducting
correlations discussed in the Introduction; 3) Work with materials
that either have weak magnetization, small domain size, or both:
As discussed previously, this approach has been used by Ryazanov
and co-workers\cite{{Ryazanov:01,Oboznov:06}} who worked with CuNi
alloy, and by Aprili and co-workers,\cite{Kontos:02} who worked
with PdNi alloy.  In CuNi alloy, Ryazanov demonstrated that the
magnetization of the CuNi makes very little contribution to the
total magnetic flux in his samples.\cite{Ryazanov:99}  He did this
by comparing the Fraunhofer pattern for a demagnetized sample with
the pattern for the same sample uniformly magnetized. The latter
pattern was shifted by a constant field, while the maximum value
of the critical current was nearly unchanged.  That shows that the
integrated vector potential stayed close to zero everywhere in the
sample; 4) Engineer the F layer to have zero net magnetic flux,
for example, by using a ``synthetic antiferromagnet." We are
currently exploring this option, and will report it in a future
publication.\cite{Mazin}

For this work, we have chosen the third option, but with PdNi
alloy rather than CuNi alloy as our weak ferromagnet.  As
discussed in the introduction, there is evidence of strong
spin-flip scattering in CuNi alloy, whereas the situation in PdNi
alloy is less clear.  It should be emphasized that the magnetism
in PdNi is quite different from the magnetism in CuNi. Because Pd
is nearly ferromagnetic itself according to the Stoner criterion,
it takes only a small concentration of Ni to make the alloy
ferromagnetic. One might expect then that the magnetism is more
uniform in PdNi than in CuNi, where magnetism results from large
Ni clusters.

As a final note regarding Nb/PdNi/Nb junctions, we found that
using PdNi layers thinner than 30nm resulted in very large
critical current densities -- so large that even our smallest
pillars ($w = 10 \mu$m) were in the regime $\lambda_J \ll w$.
Kontos \textit{et al.}\cite{Kontos:02} circumvented that problem
by introducing an additional insulating layer in their junctions
to reduce $J_c$ and hence increase $\lambda_J$.  Because our
interest is in studying junctions with large $d_F$, we omitted the
insulating layer.  This choice limited our study to junctions with
$d_F > 30$nm, which have small enough $J_c$ so that $\lambda_J
> w/4$ for our smallest diameter pillars.

\begin{figure}[tbh]
\begin{center}
\includegraphics[width=3.2in]{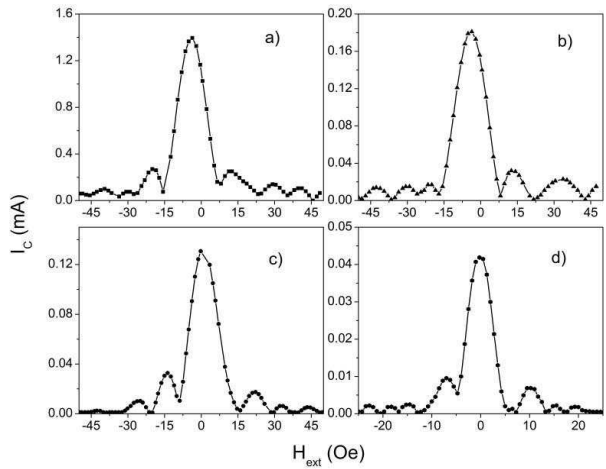}
\end{center}
\caption{Critical current vs. applied magnetic field (Fraunhofer
patterns) obtained for Nb/PdNi/Nb Josephson junctions with
different thickness of PdNi interlayer a) 35nm, b) 50nm, c) 70nm,
d) 85nm.  The pillar diameters $w$ are 10, 10, 10, and 20 $\mu$m,
respectively.} \label{Fraunhofer}
\end{figure}

\section{S/F/S Josephson Junctions with PdNi: results}
\begin{figure}[tbh]
\begin{center}
\includegraphics[width=3.2in]{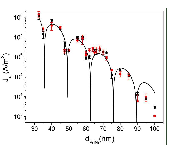}
\end{center}
\caption{(color online). Critical current density vs. $d_{PdNi}$
for all of our Nb/PdNi/Nb Josephson junctions.  Each data point
represents the average over multiple pillars on the same
substrate, with the error bars as the standard deviation of the
mean. Black points (squares) are virgin state data, while red points (circles) were
measured after magnetizing the samples (see Fig.
\ref{FraunShift}).  The solid line is a fit of Eqn.
(\ref{dirty-spin-dependent}) to the virgin state data, while
ignoring the last two data points with $d_{PdNi}=95$ and 100 nm.}
\label{Ic_vs_dF}
\end{figure}

Fig. \ref{Fraunhofer} shows $I_{c}$ vs. $H_{ext}$ data for
Nb/PdNi/Nb Josephson junctions with 35, 50, 70, and 85 nm of PdNi.
The critical current has minima whenever an integer number of flux
quanta penetrate the junction.  The extremely low values of
critical current at the minima indicate the absence of any shorts
in the insulating material surrounding the Josephson junctions.
Similar measurements were performed on samples for which the PdNi
thickness varied from 35 nm to 100 nm. The maximum current density, $J_{c}$,
obtained for all such devices is plotted vs. thickness in Fig.
\ref{Ic_vs_dF}.  This figure represents the main result of this
work.  The error bars represent the standard deviation of the mean
of the results obtained from several pillars on the same
substrate. (For the smaller values of $d_F$, we measured primarily
the pillars of diameter 10 and 20 $\mu$m, whereas for the larger
values of $d_F$, we measured the 20 and 40 $\mu$m pillars.)  The
critical current density decreases exponentially over five orders
of magnitude over this range of PdNi thickness. To our knowledge,
these data represent the widest range of ferromagnet thickness in
S/F/S Josephson junctions studied to date.

\begin{figure}[tbh]
\begin{center}
\includegraphics[width=3.2in]{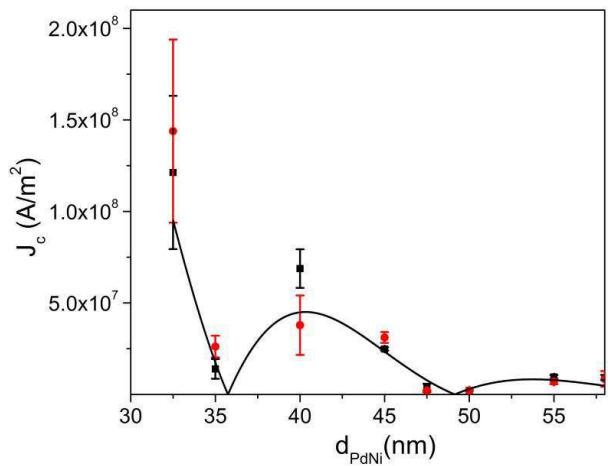}
\end{center}
\caption{(color online). Linear plot of $J_c$ vs. $d_{PdNi}$ for
the thickness range 32.5 to 58 nm.  The line is the same fit shown
in Fig. \ref{Ic_vs_dF}.} \label{LinearFit1}
\end{figure}

\begin{figure}[tbh]
\begin{center}
\includegraphics[width=3.2in]{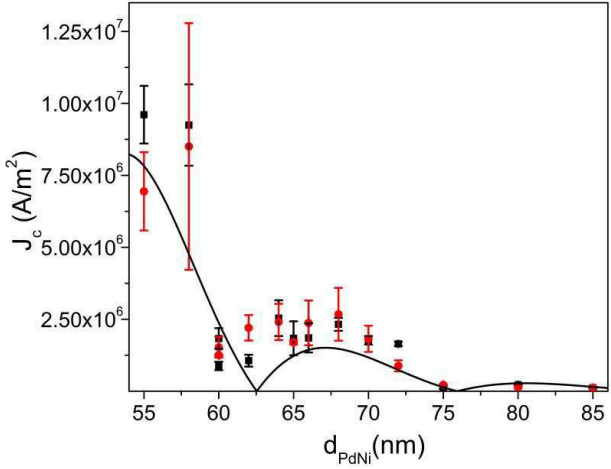}
\end{center}
\caption{(color online). Linear plot of $J_c$ vs. $d_{PdNi}$ for
the thickness range 55 to 85 nm.  The line is the same fit shown
in Fig. \ref{Ic_vs_dF}.} \label{LinearFit2}
\end{figure}

Fig. \ref{Ic_vs_dF} shows that $J_c$ does not decrease
monotonically with $d_F$, but rather exhibits local minima with
$d_F$ in the neighborhood of 35, 48, 60, and 75 nm. Figs.
\ref{LinearFit1} and \ref{LinearFit2} show $J_c$ vs $d_F$ on
linear axes, where the local minima are more clear.  Such local
minima have been observed in S/F/S junctions made with several
different ferromagnetic
materials,\cite{Oboznov:06,Kontos:02,Blum:02,Robinson:06,Sellier:03,Shelukhin:06,Weides:06,Weides:08}
and signify transitions between standard junctions and
$\pi$-junctions.

\begin{figure}[tbh]
\begin{center}
\includegraphics[width=3.2in]{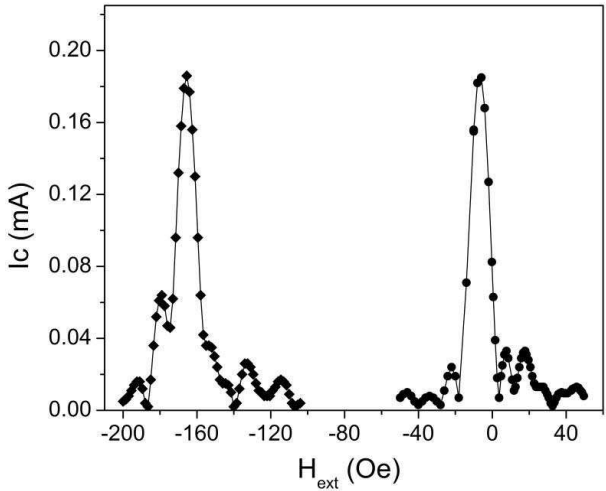}
\end{center}
\caption{Fraunhofer pattern in the virgin state (right) and after
magnetizing (left) a Nb/PdNi/Nb Josephson junction with diameter
$w=10 \mu$m and $d_{PdNi}$=47.5 nm.} \label{FraunShift}
\end{figure}

Measurements of $I_{c}$ vs. $H_{ext}$ were also performed on the
samples after magnetizing them by applying an in-plane field of 5
kOe. The resulting Fraunhofer patterns (see Fig. \ref{FraunShift})
are shifted in field to a point where the flux due to the external
field cancels out the flux due to the intrinsic magnetization. The
maximum critical currents obtained for magnetized samples match
closely with the virgin state data. This indicates that the
domains in PdNi alloy are relatively small, so that the total
excursion of the integrated vector potential as one crosses a
domain is much less than one flux quantum.

\begin{figure}[tbh]
\begin{center}
\includegraphics[width=3.2in]{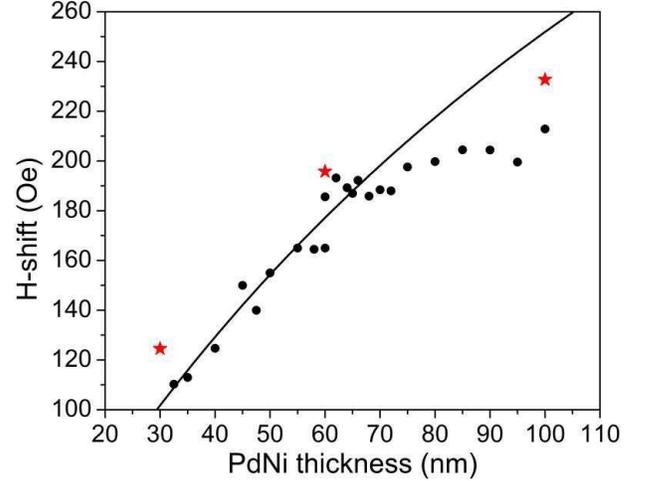}
\end{center}
\caption{(color online). Magnetic field shift of Fraunhofer
patterns of magnetized samples vs. PdNi thickness $d_F$. The line
is a fit of Eqn. (\ref{Shift}) to the data points with $d_F < 72$
nm, with $\lambda_L = 86$ nm. The fit provides an estimate of 54
emu/cm$^3$ for the remnant magnetization $M_{rem}$ of PdNi alloy.
The red stars indicate the calculated $H_{shift}$ using values of
$M_{rem}$ measured in a SQUID magnetometer on PdNi films of
thickness 30, 60, and 100 nm. (Fig. \ref{MvsH} shows the $M$ vs.
$H$ data for the 100-nm film.)}\label{Hshift}
\end{figure}

Fig. \ref{Hshift} shows the field shift of the Fraunhofer pattern
of the magnetized samples, vs. PdNi thickness $d_F$. As $d_F$
increases, the field shift saturates at a value near 200 Oe. As
the PdNi thickness increases, there is an increasing tendency for
the magnetization to rotate out of the plane, thereby decreasing
its in-plane component. The solid line is a fit to the data of
Eqn. \ref{Shift}, with the only free parameter being the remanent
magnetization $M = 54$ emu/cm$^3$. This compares with values of 76
and 62 emu/cm$^3$ measured directly on PdNi films of thickness 30
and 60 nm, respectively. The red stars in Fig. \ref{Hshift} show
the values of $H_{shift}$ calculated from Eq. \ref{Shift}, using
the values of $M_{rem}$ measured directly on PdNi films of
thickness 30, 60, and 100 nm.  The agreement with the field shifts
of the Fraunhofer patterns is reasonable.

\begin{figure}[tbh]
\begin{center}
\includegraphics[width=3.2in]{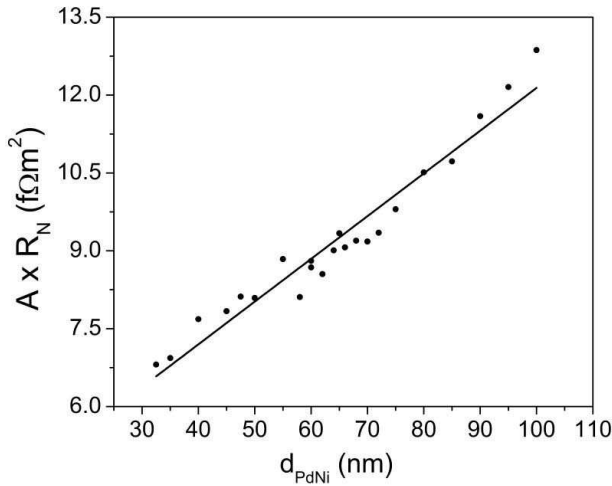}
\end{center}
\caption{Area times normal-state resistance vs. $d_{PdNi}$ for all
of our Josephson junction samples.  The slope provides the
resistivity of PdNi and the y-intercept provides twice the Nb/PdNi
boundary resistance} \label{AR_vs_dF}
\end{figure}

We have also measured the normal-state resistance of our samples
at currents much larger than $I_c$. A plot of the specific
resistance $A R_N$ (area times resistance) vs. $d_{PdNi}$ is shown
in Fig. \ref{AR_vs_dF}. The interface and bulk contributions to $A
R_{N}$ are given by:
\begin{equation}\label{int}
A R_{N} = 2 A R_{B} + \rho_{PdNi} d_{F}
\end{equation}
where $\rho_{PdNi}$ is the resistivity of PdNi, $d_{F}$ is the
thickness of the PdNi layer, and $R_{B}$ is the Nb/PdNi boundary
resistance.  A linear fit to all of the data gives a boundary
resistance of $A R_{B} = 1.95 \pm 0.14 f\Omega m^2$ and a
resistivity of PdNi, $\rho_{PdNi}=82 \pm 4 n\Omega m$.  There is
some indication in the data that the slope increases for $d_{F}$
greater than about 75 nm.  If we fit only the data for $d_F < 75$
nm, then we find $A R_{B} = 2.39 \pm 0.14 \Omega m^2$ and
$\rho_{PdNi}= 62 \pm 5 n\Omega m$.  Independent measurements of
the in-plane PdNi resistivity were performed on 200-nm thick
films, using the van der Pauw method. Those measurements produced
the value $\rho_{PdNi}= 116 \pm 2 n\Omega m$.  It is plausible
that the in-plane resistivity is larger than the perpendicular
resistivity if the PdNi films grow in a columnar fashion, although
measurements on other sputtered metals often find quite close
agreement between these two measurement methods.\cite{BassPratt}

\section{Theory of S/F/S Josephson Junctions}

There have been a large number of theoretical works dealing with
S/F/S Josephson junctions.  There are three energy scales whose
relative size determines three distinct regimes. The energy scales
are the exchange energy in the ferromagnet, $E_{ex}$, the gap in
the superconductor, $\Delta$, and $\hbar / \tau$, the inverse of
the mean free time between collisions of an electron propagating
in the ferromagnet.  In all of the experimental work on S/F/S
Josephson junctions published to date, including this work,
$E_{ex} \gg \Delta$.  There is a wide variation, however, in the
size of $\hbar / \tau$ relative to those two energies.  The true
clean limit is expressed by $E_{ex} \tau \gg \hbar$, which also
implies $\Delta \tau \gg \hbar$. The intermediate limit is where
$E_{ex} \tau \gg \hbar$ but $\Delta \tau \ll \hbar$, whereas the true
dirty limit is where both $\Delta \tau \ll \hbar$ and $E_{ex} \tau
\ll \hbar$.  These three regimes can also be characterized by the
relative sizes of the three length scales: the electron mean free
path, $l_e = v_F \tau$, the superconducting coherence length,
$\xi_S = \hbar D_S/\Delta$, and the clean-limit exchange length
discussed earlier, $\xi_F = \hbar v_F/2E_{ex}$.

The simplest limit is the dirty limit, with the additional
constraint that the ferromagnetism is weak enough so that one can
treat the spin-up and spin-down bands identically, i.e. as having
the same Fermi velocity and mean free path.  In this limit, the
Usadel equation is valid.  An expression for the critical current
$I_c$ as a function of ferromagnetic layer thickness $d_F$ was
first derived by Buzdin \textit{et al.}\cite{Buzdin:92} In this
regime, the oscillation and decay of $I_c$ as a function of $d_F$
are both governed by a single length scale -- the "dirty-limit"
exchange length, $\xi_F^* = (\hbar D_F/E_{ex})^{1/2}$. Once $d_F$
exceeds $\xi_F^*$, the thickness dependence of $I_c$ takes the
simple form:
\begin{equation}\label{dirty-simple}
I_c(d_F) =
I_{c0}*\exp(\frac{-d_F}{\xi_F^*})*|\sin(\frac{d_F}{\xi_f^*} + \pi
/4)|
\end{equation}
In the presence of spin-flip or spin-orbit scattering, Eqn
(\ref{dirty-simple}) is modified, and the length scales governing
the decay and the oscillation are no longer
equal.\cite{Faure:06,Bergeret:03-2,Demler,Bergeret:07} The more
general form can be written as
\begin{equation}\label{dirty-spin-dependent}
I_c(d_F) =
I_{c0}*\exp(\frac{-d_F}{\xi_{F1}})*|\sin(\frac{d_F}{\xi_{F2}} +
\phi)|
\end{equation}
In general, the effect of spin-flip or spin-orbit scattering is to
shorten the decay length scale, $\xi_{F1}$, relative to $\xi_F^*$,
and to lengthen the oscillation length scale, $\xi_{F2}$.  In the
presence of sufficiently strong spin-orbit (but not spin-flip)
scattering, the oscillations disappear completely. An equation
similar to Eqn. (\ref{dirty-spin-dependent}) has successfully been
used to fit $I_c$ vs. $d_F$ data from S/F/S junctions containing
CuNi alloy, with $\xi_{F1} = 1.3$nm, and $\xi_{F2} =
3.5$nm.\cite{Oboznov:06} The very short value of $\xi_{F1}$
compared to $\xi_{F2}$ was interpreted as implying that spin-flip
scattering is strong in that material.

Eqn. (\ref{dirty-spin-dependent}) can also be fit to our data, as
shown in Figs. \ref{Ic_vs_dF}-\ref{LinearFit2}.  But in our case,
the length scale governing the exponential decay ($\xi_{F1} =
8.0\pm 0.5$nm) is considerably longer than the length scale
governing the oscillation ($\xi_{F2} =  4.1\pm 0.1$nm), hence the
dirty-limit hypotheses that led to Eqn.
(\ref{dirty-spin-dependent}) are not fulfilled. However, the
condition $\xi_{F1} > \xi_{F2}$ has also observed in S/F/S
Josephson junctions containing the strong ferromagnets: Ni, Fe,
Co, and Ni$_{80}$Fe$_{20}$ (also known as Permalloy, or
Py).\cite{Robinson:06}  Those materials have very large exchange
energy, hence they are in the intermediate limit, with $E_{ex}
\tau \gg \hbar$ but still $\Delta \tau \ll \hbar$. Our observation
of $\xi_{F1} > \xi_{F2}$ in S/F/S junctions with PdNi alloy
suggest that PdNi may also be in the intermediate limit.

The intermediate limit has been studied in several theoretical
papers. Bergeret \textit{et al.}\cite{Bergeret:01} provide a
general formula for the critical current, valid both in the dirty
limit and intermediate limit.  In the intermediate limit the
formula simplifies when the F-layer thickness is larger than the
mean free path.  Eqn. (20) from ref. \onlinecite{Bergeret:01} is:
\begin{equation}\label{intermediate-Bergeret}
I_c(d_F) \propto \pi T
\sum_{\omega>0}\frac{\Delta^2}{\Delta^2+\omega^2}\frac{\sin(d_F/\xi_F)}{d_F/\xi_F}\exp(\frac{-d_F}{l_e}(1+2\omega\tau))
\end{equation}
where the sum is over the positive Matsubara frequencies,
$\omega_m = \pi k_B T (2m+1)$ with $T$ the temperature.
Asymptotically at large values of $d_F$, Eqn.
(\ref{intermediate-Bergeret}) is quite similar to Eqn.
(\ref{dirty-spin-dependent}), with $\xi_{F1} = l_e$ and $\xi_{F2}
= \xi_F$.  Nevertheless, we have fit Eqn.
(\ref{intermediate-Bergeret}) directly to our data, and obtained
the parameters $\xi_F = 4.0 \pm 0.1$nm and $l_e = 11.2 \pm 1$nm.
Not surprisingly, the former agrees closely with the value of
$\xi_{F2}$ obtained from the fit of Eqn.
(\ref{dirty-spin-dependent}), while the value of $l_e$ obtained
from Eqn. (\ref{intermediate-Bergeret}) is somewhat larger than
the value of $\xi_{F1}$ obtained from the fit of Eqn.
(\ref{dirty-spin-dependent}) -- the difference undoubtedly due to
the sum over Matsubara frequencies in Eqn.
(\ref{intermediate-Bergeret}).  We will use the larger value of
$l_e$ in the following discussion.

An alternative theoretical approach was taken by Kashuba, Blanter,
and Fal'ko,\cite{Kashuba} who considered a model of S/F/S
junctions taking explicit account of spin-dependent and spin-flip
scattering in the F layer. In the intermediate limit, these
authors find a result that is nearly identical to Eqn.
(\ref{intermediate-Bergeret}).

\section{Discussion of Results}
\subsection{Estimate of mean free path and exchange energy in PdNi}

In many metals, it is straightforward to estimate the mean free
path directly from the measured resistivity, as the product $\rho
l_e$ is inversely proportional to the Fermi surface area, and has
been tabulated for a large number of metals.\cite{BassLandolt} In
PdNi alloy, however, the $\rho l_e$ product is not known.  Some
workers have tried to estimate $\rho l_e$ in PdNi from its value
in Pd, but even that estimation is not straightforward, due to the
complex band structure of Pd. To illustrate the difficulty,
previous workers have quoted values of the $\rho l_e$ product as
small as $0.33 f \Omega m^2$ (refs. \onlinecite{Cirillo,Matsuda})
and as large as $4.0 f \Omega m^2$ (ref.
\onlinecite{Kontos-thesis}). The former value is certainly too
small, because it was calculated using the Einstein relation,
$\sigma = n(E_F) e^2 D$, with the density of states at the Fermi
level, $n(E_F)$, obtained from the electronic specific heat
coefficient $\gamma = (\pi^2/3)k_B^2 n(E_F)$.  The problem is
that, in Pd, the specific heat is dominated by heavy holes on the
open ``jungle gym" portion of the Fermi surface, while the
electronic transport is dominated by much lighter electrons on a
part of the surface centered at the $\Gamma$-point.  (It is common
to refer to these electrons as ``s-like"; but that is incorrect
because they are strongly hybridized with the d
bands.\cite{Pinski}) Discussion of the Fermi surface in Pd has
been given in several
papers.\cite{Pinski,Vuillemin,Dye,Mazin:1984} Pinski \textit{et
al.}\cite{Pinski} state that the $\Gamma$-centered sheet of the
Fermi surface carries the vast majority of the transport current
-- up to 97\% at 10 K.

We estimate the $\rho l_e$ product for PdNi in two ways.  In his
Ph.D. thesis,\cite{Kontos-thesis} Kontos found that the
resistivity of thin PdNi films with $x \approx 12 \%$ varied
linearly with inverse thickness $1/d$ once the films were thinner
than about 8 nm, indicating that the mean free path is limited by
the film thickness.  The slope of the graph gives the product
$\rho d = 1.7 f \Omega m^2$. We expect the $\rho l_e$ product to
be within a factor of two of this value. The second method relies
on the statement by Pinksi \textit{et al.}\cite{Pinski} that
electronic transport in Pd is dominated by the electrons on the
$\Gamma$-centered sheet. The total number of carriers in that
band, the density of states at the Fermi level, the effective
mass, and the Fermi velocity on that sheet have all been tabulated
by Dye \textit{et al.}\cite{Dye} based on deHaas van-Alphen
measurements of the Fermi surface of Pd. The values of those four
quantities are: $n = 0.375$ carriers/atom $= 2.54 \cdot 10^{28}
m^{-3}$; $n(E_F) = 0.189$ states/(eV atom spin) $=2.56
\cdot10^{28} eV^{-1}m^{-3}$; $m^* = 2.0 m_e$; and $v_F = 0.6 \cdot
(2\pi/a)\cdot (\hbar/2m) = 5.6 \cdot10^5$m/s.\cite{DumoulinNote}
Using the Drude formula, $\sigma = n e^2 \tau / m^*$, one finds
$\rho l_e = 1.55 f \Omega m^2$, while using the Einstein relation
one finds $\rho l_e = 1.31 f \Omega m^2$. (The slight difference
between the two values is likely due to the non-parabolic
character of the $\Gamma$-centered sheet.)  These values are close
to the value $\rho l_e = 1.7 f \Omega m^2$ estimated from the
thickness dependence found by Kontos.  If we take our own measured
resistivity, $\rho = 82 n \Omega m$, and the value of the mean
free path from the $J_c$ vs. $d_F$ fit, $l_e = 11.2$nm, we obtain
$\rho l_e = 1.0 f \Omega m^2$, which is not too far from the
estimates given above.  (The only mystery we can not explain from
this analysis is the very low value of the Fermi velocity, $v_F =
2.0 \cdot 10^5$m/s, measured by Dumoulin \textit{et
al.}\cite{Dumoulin} in a proximity effect experiment between Pb
and Pd.  It is unclear why that experiment measures the low Fermi
velocity of the open hole sheet rather than the higher Fermi
velocity of the $\Gamma$-centered sheet.)

\begin{table*}[tbh]
\begin{tabular}
[c]{|c|c|c|c|c|c|c|c|c|c|c|c|}\hline
Source & Experiment & Ni conc. & $v_F$ & $\rho l_e$ & $\rho$ & $l_e$ & $\xi_{F1}$ & $\xi_{F2}$ & $E_{ex}$ & formula to & $T_{Curie}$ \\
& & (at. \%) & ($10^5$m/s) & (f$\Omega$m$^2$) & (n$\Omega$m) & (nm)
& (nm) & (nm) & (meV) & extract $E_{ex}$ & (K) \\\hline
\onlinecite{Kontos:01} & S/F DOS & 10 & 2.0 & 4 & - & $\approx d_F$ & 2.8 & 2.8 & 35 & $(\hbar D/E_{ex})^{1/2}$ & 100 \\
\onlinecite{Kontos:02} & S/I/F/S & 12 & 2.0 & 4 & - & $\approx d_F$ & 3.5 & 3.5 & 13 & $(\hbar D/E_{ex})^{1/2}$ & 260 \\
\onlinecite{Cirillo} & $T_{c}$ of S/F & 14 & 2.0 & 4 & 240 & 16.6 & 6 & 6 & 15 & $(\hbar D/E_{ex})^{1/2}$ & 156\\
\onlinecite{Cirillo:02,Cirillo:03} & $T_{c}$ of S/F & 14 & 2.0 & 0.96 & 240 & 4 & 3.4 & 3.4 & 13 & $(\hbar D/E_{ex})^{1/2}$ & 185 \\
\onlinecite{Matsuda} & $T_{c}$ of S/F & 15 & 2.0 & 0.3256 & 220 & 1.48 & 3.5 & 3.5 & 13 & $(\hbar D/E_{ex})^{1/2}$ & 260 \\
\onlinecite{Bauer} & S/F/S & 18 & 2.0 & - & - & - & 2.4 & 2.4 & 52 &
$(\hbar D/E_{ex})^{1/2}$ & 200 \\
this work & S/F/S & 12 & 5.6 & 1.0 & 82 & 11 & 8.0 & 4.0 & 44 & $\hbar
v_F/E_{ex}$ & 175
\\\hline
\end{tabular}
\newline \caption{PdNi parameters from several groups.
The superconductor used in all the above experiments was Niobium.
Note that most workers have used the diffusive formula, $\xi_{F2}
= (\hbar D/E_{ex})^{1/2}$, to extract $E_{ex}$ from the measured
value of $\xi_{F2}$, whereas we have used the ballistic formula,
$\xi_{F2} = \hbar v_F/2E_{ex}$.  Our choice of Fermi velocity,
$v_F$, is discussed in the text and in ref.
\onlinecite{DumoulinNote}.} \label{tableParam}
\end{table*}

From the period of oscillation of $J_c$ vs. $d_F$, we found
$\xi_{F2} = 4.0 \pm 0.1$nm.  Using the Fermi velocity of the
dominant carriers, $v_F = 5.6 \cdot10^5$m/s, gives an estimate for
the exchange energy in our Pd$_{88}$Ni$_{12}$ alloy of $E_{ex} =
\hbar v_F / 2 \xi_{F2} = 44$meV.  This value is somewhat higher
than values quoted by previous workers
\cite{Kontos:02,Cirillo,Cirillo:02,Cirillo:03,Matsuda}, but those
earlier estimates were either made using the smaller value of
$v_F$, or using the diffusive formula $\xi_F^*=\sqrt(\hbar
D/E_{ex})$.  A more meaningful comparison is of the length scales
$\xi_{F1}$ and $\xi_{F2}$ found in different experiments. For
example, Kontos \textit{et al.}\cite{Kontos:02} found $\xi_{F1}
\approx \xi_{F2} = 2.8 nm$ in their study of S/I/F/S Josephson
junctions with a PdNi alloy of similar concentration to ours.  The
values of $\xi_{F2}$ in their experiment and ours are rather close
to each other, but the values of $\xi_{F1}$ are not.  It is not
clear if that discrepancy is significant or not.  The thickness
range covered in the earlier work was 4.5 - 14 nm, whereas the
range we covered was 32.5 - 100 nm.  If the mean free path in the
PdNi alloy is indeed in the range of 11 nm, then the samples
studied by Kontos \textit{et al.} were in the crossover regime
with $d_F \approx l_e$, where the thickness dependence has not yet
obtained the asymptotic exponential decay, $J_c \propto
\exp(-d_F/l_e)$. But then one would expect a less steep decay of
$J_c$ with $d_F$, rather than a more steep decay.  Perhaps a more
relevant observation is simply that the PdNi films deposited in
different laboratories may have different polycrystalline
structures, and hence very different mean free paths.  A summary
of the parameters estimated by previous workers, as well as by our
work, is given in Table I.

\subsection{Spin-triplet superconducting correlations?}

One of the primary goals of this work was to search for signs of
spin-triplet superconducting correlations in our samples. At first
glance, the data in Fig. (\ref{Ic_vs_dF}) show no sign of
spin-triplet superconducting correlations, which might manifest
themselves as a crossover to a less-steep exponential decay of
$J_c$ at large values of $d_F$.  It is intriguing, however, that
the length scale characterizing the exponential decay of $J_c$ in
our samples, $\xi_{F1} = 8 - 11$nm, is substantially longer than
that observed previously in shorter S/F/S junctions with
PdNi.\cite{Kontos:02}  Could it be that we are already observing
the triplet Josephson effect throughout the whole range of $d_F$
reported here?  The strongest evidence against such an
interpretation is the nearly-periodic set of local minima we
observe in $J_c$ vs. $d_F$, shown in Figures
(\ref{Ic_vs_dF}-\ref{LinearFit2}).  We believe that those local
minima signal crossovers between 0-junctions and $\pi$-junctions,
which are due to the effect of the exchange splitting on
spin-singlet superconducting correlations.  A Josephson
supercurrent dominated by spin-triplet correlations would not
exhibit such minima, but rather would decay monotonically with
increasing $d_F$.  Nevertheless, to rule out the spin-triplet
hypothesis definitively would require stronger evidence that the
local minima we observe truly represent $0-\pi$ crossovers, rather
than an unlucky distribution of uncertainties in the data that
mimics a periodic set of local minima.

There are three ways one could make a more stringent test that the
local minima observed in our $J_c$ vs. $d_F$ data are due to
$0-\pi$ crossovers: 1) A direct measurement of the current-phase
relationship of the junction;\cite{Frolov:2004} 2) Extension of
$J_c$ vs. $d_F$ measurements to smaller values of $d_F$, to see if
the slope of the $J_c$ vs. $d_F$ semi-log plot changes to a value
close to that measured by Kontos \textit{et al.}.  This would
require reducing the lateral size $w$ of our junctions, so as to
maintain the condition $\lambda_J > w/4$;  3) Measurement of $J_c$
vs. temperature $T$ for samples very close to a $0-\pi$ crossover.
In S/F/S junctions with very weak ferromagnets, the $0-\pi$
crossover has been observed in the $T$-dependence of
$J_c$.\cite{Oboznov:06,Ryazanov:01,Sellier:03,Weides:08} As the exchange energy increases, however,
the thickness range over which one can see a non-monotonic
$T$-dependence of $J_c$ gets progressively
narrower.\cite{Weides:08} Each of these checks presents its own
set of challenges, and represents a possible direction for future
work.

\subsection{Spin-flip scattering in PdNi}

Assuming that our data are not the result of spin-triplet
correlations, we would like to know why not. There are several
factors that may contribute.  First and foremost, strong spin-flip
scattering, if it exists in our PdNi alloy, would destroy the
triplet correlations.  To address this issue, we have
independently tried to measure directly the spin diffusion length,
$l_{sf}$, in PdNi alloy using techniques borrowed from the Giant
Magnetoresistance (GMR) community.  A complete discussion of those
measurements is given elsewhere.\cite{Hamood} Here we note the
most salient results.  First, the value of the spin diffusion
length obtained, $l_{sf} = 2.8 nm$, is extremely surprising given
the much longer length scale characterizing the decay of the
Josephson supercurrent in the present work. Normally, one assumes
that spin-flip and spin-orbit scattering processes occur on length
scales much longer than the mean free path, justifying the
diffusive model used in discussing the spin diffusion
length.\cite{ValetFert:93} Hence a measurement of a spin diffusion
length several times shorter than the mean free path is difficult
to interpret.  We believe that the very short value of $l_{sf}$
observed using a GMR spin valve\cite{Hamood} may be due to the
out-of-plane magnetic anisotropy of PdNi discussed earlier.  The
analysis of the spin-valve data of ref. \onlinecite{Hamood} assume
that the magnetizations of the two magnetic layers are either
parallel or antiparallel to each other. Our recent discovery of
the out-of-plane magnetic anisotropy in PdNi casts doubt on this
assumption.  It is possible that the magnetization of PdNi is
inhomogeneous on very short length scales, which may then destroy
spin memory by rotating the spin on a length scale smaller than
the mean free path. Clarifying this issue will require further
experiments.

A second possible reason we do not observe signs of spin-triplet
superconducting correlations is that the length scale
characterizing the magnetic inhomogeneity in PdNi might not be
comparable to the Cooper pair coherence length $\xi_s$ in Nb.  Let
us refer to the length scale characterizing the magnetic
inhomogeneity as $\delta_m$. If $\delta_m \ll \xi_s$, then a
Cooper pair will experience the magnetization averaged over the
length $\xi_s$, and triplet correlations will not be produced
efficiently.  If $\delta_m \gg \xi_s$, then a typical Cooper pair
experiences no magnetic inhomogeneity. The coherence length in our
Nb is known to be about 13 nm.\cite{Gu:02} Unfortunately, the
typical domain sizes and domain wall widths in our PdNi alloy are
not known. Because the Curie temperature of PdNi is well below
room temperature, obtaining information about magnetic structure
requires a low temperature magnetic visualization technique, such
as low-temperature magnetic force microscopy (MFM) or Bitter
decoration. And even then, the former method is not well-suited to
weak ferromagnets, because the magnetization of the MFM tip may
influence the domain structure of the sample.  Very recently the
domain structure of CuNi alloy was measured using the Bitter
decoration technique.\cite{Veshchunov} Those measurements
confirmed the out-of-plane magnetic anisotropy of that material,
and found a typical domain size of 100nm in the virgin state or at
the coercive field.  Unfortunately, no such measurements have been
performed on PdNi alloy, to our knowledge. Clearly, a thorough
study of the magnetic domain structure of PdNi would help clarify
this issue.

\section{Future Directions}

The most urgent work needed in the future is a strong verification
(or repudiation) that the local minima in our data do indeed
represent $0-\pi$ crossovers, rather than sample-to-sample
fluctuations.  Looking further ahead, to have any hope of nailing
down the elusive spin-triplet supercurrent in S/F/S junctions will
require better characterization of magnetic materials. On the one
hand, the spin diffusion length is a crucial parameter, as it
limits the spatial extent of spin-triplet correlations.
Fortunately, the spin-diffusion length has been measured in some
ferromagnetic materials,\cite{BassPrattReview} but more work is
needed.  Of equal important is information about the typical
length scales characterizing the domain structure of ferromagnetic
thin films. This is a complex issue, as either the domain size or
domain wall width may be important.  For example, in a film where
neighboring domains have antiparallel magnetization, the
long-range triplet component is generated only in the domain
walls,\cite{Fominov:07} hence it is crucial that the domain wall
width be comparable to $\xi_s$. If, however, neighboring domains
have non-collinear magnetizations, then it would seem that the
long-range triplet could be produced even if the domain walls are
very thin, as long as the typical domain size is comparable to
$\xi_s$. Unfortunately, domain sizes in polycrystalline films are
not known \textit{a priori}.\cite{OHandleyBook}

\section{Conclusions}

We have measured the critical current of Nb/PdNi/Nb Josephson
junctions with PdNi thicknesses ranging from 32.5 to 100 nm. The
critical current drops by five orders of magnitude over this
thickness range.  The data appear to have a periodic array of
local minima, suggesting that the supercurrent is of the
conventional spin-singlet type over the entire thickness range. We
have discussed possible reasons for the absence of induced
spin-triplet correlations, such as spin-flip scattering or a poor
match of length scales between the magnetic domains and the
superconducting coherence length.

\section{Acknowledgements}

We are grateful to M. Aprili, Y. Blanter, T. Kontos, and V.
Ryazanov for helpful discussions, and to B. Bi, R. Loloee, and Y.
Wang for technical assistance. This work was supported by the
Department of Energy under grant ER-46341.

\end{document}